\documentclass[12pt]{article}

\usepackage{amsmath,amssymb,amsthm,amscd,mathtools, graphicx, mathtext,color,wrapfig}

\usepackage[numbers,sort&compress]{natbib}
\makeatletter

\@addtoreset{equation}{section}
\makeatother

\newcommand{\beq}{\begin{equation}}
\newcommand{\eeq}{\end{equation}}

\theoremstyle{definition}

\begin{document}
\baselineskip=18pt  
\baselineskip 0.7cm
\begin{titlepage}

\setcounter{page}{0}

\renewcommand{\thefootnote}{\fnsymbol{footnote}}

\begin{flushright}
ITEP-TH 30/13\\
SISSA 42/2013/FISI
\end{flushright}

\vskip 1.0cm

\begin{center}
{\LARGE \bf
Gauge/Liouville Triality
\\
\vskip  0.5cm
}

\vskip 0.5cm

{\large
Mina Aganagic$^{1,2}$, Nathan Haouzi$^{1}$,
Can Koz\c{c}az$^{3,4}$, \\
and Shamil Shakirov$^{1,2,5}$}
\\
\medskip

\vskip 0.5cm

{\it
$^1$Center for Theoretical Physics, University of California, Berkeley,  USA\\
$^2$Department of Mathematics, University of California, Berkeley,  USA\\
$^3$International School of Advanced Studies (SISSA), Trieste, Italy \\
$^4$Instituto Nazionale di Fizika Nucleare, Trieste, Italy \\
$^5$Institute for Theoretical and Experimental Physics, Moscow, Russia
}

\end{center}

\vskip 0.5cm

\centerline{{\bf Abstract}}
\medskip

Conformal blocks of Liouville theory have a Coulomb-gas representation as Dotsenko-Fateev (DF) integrals over the positions of screening charges. For $q$-deformed Liouville, the conformal blocks on a sphere with an arbitrary number of punctures are manifestly the same, when written in DF representation, as the partition functions of a class of 3d  $U(N)$ gauge theories with ${\cal N}=2$ supersymmetry, in the $\Omega$-background. Coupling the 3d gauge theory to a flavor in fundamental representation corresponds to inserting a Liouville vertex operator; the two real mass parameters determine the momentum and position of the puncture. The DF integrals can be computed by residues. The result is the instanton sum of a five dimensional ${\cal N}=1$ gauge theory.
The positions of the poles are labeled by tuples of partitions, the residues of the integrand are the Nekrasov summands.

\noindent\end{titlepage}
\setcounter{page}{1} 


\section{Introduction }

A useful tool for studying Liouville theory is the Dotsenko-Fateev (DF) free-field representation of the Liouville conformal blocks \cite{Dotsenko:1984nm, Dijkgraaf:2009pc, Schiappa:2009cc, Mironov:2009ib, Mironov:2010zs, Mironov:2010su, Mironov:2010pi, Awata:2010yy, Zhang:2011au}. We will show that the Dotsenko-Fateev integrals of a $q$-deformed Liouville conformal field theory \cite{Shiraishi:1995rp, Awata:1996xt, Awata:2010yy} have a physical interpretation.  They are partition functions of a 3d ${\cal N}=2$ gauge theory, which we will call ${\cal G}_{3d}$, in the 3d $\Omega$-background $M_q=({\mathbb C }\times S^1)_q$. For conformal blocks on a sphere with $M+2$ punctures, ${\cal G}_{3d}$ is a $U(N)$ gauge theory with $M$ flavors in the fundamental representation. The equivalence of the $q$-Liouville conformal block and the gauge theory partition function is manifest. The screening charge integrals of DF are the integrals over the Coulomb branch of the gauge theory. Inserting the Liouville vertex operators corresponds to coupling the 3d gauge theory to a flavor. The momentum and position of the puncture are given by the real masses of the two chirals within the flavor. 

The DF formulation of Liouville conformal blocks was considered just a tool to study Liouville, and an imperfect one, because of the constraint that some of the momenta be integral. Now, it is apparent why the constraint on the momenta is needed: in the 3d gauge theory, to compute the partition function on $M_q$ one has to choose a vacuum, and the integer momenta are ranks of gauge groups that are left unbroken by this choice.

The correspondence between $q$-Liouville CFT and the 3d gauge theory ${\cal G}_{3d}$ has another leg.
The 3d theory ${\cal G}_{3d}$ is a theory on vortices in a gauge theory ${\cal G}_{5d}$ in two higher dimensions. The 5d gauge theory has ${\cal N}=1$ supersymmetry, $U(M)$ gauge symmetry and $2M$ fundamental hypermultiplets. The theory on charge $N$ vortices in ${\cal G}_{5d}$, worked out in \cite{HananyTong, HananyTong2}, is the theory we called ${\cal G}_{3d}$.

The $q$-Liouville conformal block and the partition function of ${\cal G}_{3d}$ have a direct interpretation in terms of ${\cal G}_{5d}$ as well. The partition function of ${\cal G}_{3d}$ can be computed by residues. The sum over the residues is the instanton sum of the 5d gauge theory ${\cal G}_{5d}$ on the 5d $\Omega$-background $M_{q,t}=({\mathbb C }\times {\mathbb C}\times S^1)_{q,t}$, restricted to the locus of the moduli space where Coulomb branch parameters take integer values. The positions of the poles are labeled by tuples of partitions, and the integrands are equal to Nekrasov summands.

The most surprising aspect of the duality is that it can be made manifest. The equality of the partition functions of ${\cal G}_{3d}$, ${\cal G}_{5d}$ and the conformal blocks of $q$-Liouville can be proven by inspection. The relation between the three holds at any $N$. If one wants to obtain Liouville correlators at arbitrary, continuous momenta, one should take the large $N$ limit. This is also necessary if one wishes to go beyond conformal blocks to Liouville partition functions.

Thus, we find a triality of correspondences relating ${\cal G}_{3d}$,  ${\cal G}_{5d}$, and $q$-Liouville.  The relation between the 5d gauge theory and $q$-Liouville is reminiscent of the AGT correspondence \cite{AGT} in one dimension higher. There is a difference, though: ${\cal G}_{5d}$ is not the gauge theory AGT would have naturally associated to Liouville; rather, it is its spectral dual \cite{Mironov:2012uh, Mironov:2013xva} theory ${\tilde {\cal G}}_{5d}$. ${\cal G}_{5d}$ and ${\tilde{\cal G}}_{5d}$ share the same spectral curve and have the same M-theory origin, but a different Lagrangian. The AGT correspondence is a composite of the duality we find here, and the spectral duality.

The paper is organized as follows. In section 2, we study the partition function ${\cal Z}_{{\cal G}_{3d}}$ of the 3d theory ${\cal G}_{3d}$ on $M_q$. In section 3, we recall the DF representation of the $q$-Liouville confomal blocks on a sphere, and show they manifestly equal ${\cal Z}_{{\cal G}_{3d}}$. In section 4, we study the Nekrasov partition function ${\cal Z}_{{\cal G}_{5d}}$ of the 5d gauge theory ${\cal G}_{5d}$ on $M_{q,t}$. We prove that, for quantized values of Coulomb branch moduli, ${\cal Z}_{{\cal G}_{5d}}$ coincides with ${\cal Z}_{{\cal G}_{3d}}$ for any number of punctures. In section 5, we explain the geometry behind this correspondence. In section 6, we discuss the physical reason for the triality and the relation to the work of \cite{AGT}.

\section{The 3d Gauge Theory ${\cal G}_{3d}$}

The theory ${\cal G}_{3d}$ is obtained by starting with ${\cal N}=4$  $U(N)$ gauge theory. The theory has $SU(2)_L\times SU(2)_R$ R-symmetry.  The 3 vector multiplet scalars are neutral under $SU(2)_R$ and transform in the spin $1$ representation of $SU(2)_L$.  From the ${\cal N}=2$ perspective, the theory has two $U(1)$ symmetries, which are the diagonal subgroups of $SU(2)_R$ and $SU(2)_L$. The off-diagonal subgroup of the two $SU(2)$'s is a flavor symmetry, $U(1)_{t}$. We break supersymmetry to ${\cal N}=2$ by turning on a real mass $\epsilon$ corresponding to it \cite{Ooguri:1999bv,Aganagic:2000gs,Hellerman:2011mv}. This gives the adjoint chiral from the ${\cal N}=4$ vector multiplet has mass $-\epsilon$.

We will couple this theory to $M$ flavors in fundamental representation. Each flavor has the matter content of the ${\cal N}=4$ hypermultiplet -- it consists of a pair of ${\cal N}=2$ chiral multiplets in fundamental and anti-fundamental representation -- but the superpotential coupling to the adjoint in the vector multiplet, implied by ${\cal N}=4$ supersymmetry, is absent. We can give such hypermultiplets arbitrary real masses -- there are $2M$ parameters in all.\footnote{A  similar theory was studied in \cite{GK}, in connection to integrable systems, but in this case the theory had honest ${\cal N}=4$ supersymmetry before $\epsilon$ deformation.}. 

We compactify ${\cal G}_{3d}$ on the 3d $\Omega$ background:

$$
M_q = ({\mathbb C} \times S^1)_q.
$$
As we go around the $S^1$ we simultaneously rotate the complex plane by $q$ and twist by the $U(1)_R$-symmetry, to preserve supersymmetry. The partition function of the theory on $M_q$ computes the index

\beq\label{3dtrace}
{\cal Z}_{{\cal G}_{3d}}={\rm Tr} (-1)^F q^{S-S_R} t^{S_R-S_L},
\eeq
where $S$ is the generator of rotations, and $S_R$ and $S_L$ are the $U(1)$ generators in $SU(2)_L\times SU(2)_R$. We denoted
$t=e^{\epsilon}$, and let $q=e^{\hbar}$.

\beq\label{part}
{\cal Z}_{{\cal G}_{3d}}={1\over N!}\int d^Nx \;\;  \Phi_{V}(x)\,  \prod_{a=1}^{M} \Phi_{H_a}(x)  \; e^{k \,{\rm Tr}x^2/2\hbar + \zeta \,{\rm Tr} x/\hbar} .
\eeq
Here, $x$'s are Coulomb branch parameters -- since the theory is on a circle, they are complex.
The integrand is a product including all contributions of the massive BPS particles in the theory, the $W$ bosons, $\Phi$, and any additional matter.
The exponent  contains the classical terms: the FI parameter $\zeta$, and the Chern-Simons level $k$. The Chern-Simons level is absent in the ${\cal N}=4$ theory, but allowed in ${\cal N}=2$. If the gauge symmetry were just a global symmetry, $x$'s would have been parameters of the theory and the partition function of the theory would have been the integrand. Gauging the $U(N)$ symmetry corresponds to simply integrating\footnote{This partition function is the index studied in \cite{AS, AS2, Aganagic:2012au} with application to knot theory; see also \cite{Fuji:2012nx}. The index is a chiral building block of the $S^3$ or $S^2\times S^1$ partition functions \cite{Hama:2010av,Kapustin:2011jm, Hama:2011ea,Pasquetti:2011fj, Nieri:2013yra, BDP, Taki:2013opa}, deformed by $t$, the fugacity of a very particular flavor symmetry. Given the flavor symmetry, to get the block from the $S^3_b$ partition function, say, it suffices to pick out the terms which are perturbative in $b$. This replaces the non-compact quantum dilogarithm $s_b$ by its Faddeev-Kashaev version \cite{Faddeev:1993rs} $\varphi(x)$ we use here.} over $x$.

The contributions are as follows: the contribution of the hypermultiplet in the fundamental representation is

\beq\label{basic}
\Phi_{H}(x)= \prod_{1\leq i \leq N}  \prod_{n=0}^{\infty}{ 1 -q^n u \; e^{x_i-m_-}\over 1-q^n v \; e^{x_i-m_+}},
\eeq
where $u = (qt)^{1/2}$ and $v=(q/t)^{1/2}$. The right hand side is written in terms of Faddeev-Kashaev quantum dilogarithms.
We can think of the flavor in the fundamental representation in one of two equivalent ways: it as a pair of ${\cal N}=2$ chiral multiplets, one in the fundamental and the other in the anti-fundamental representation, which then have the same $U(1)_t$ charge. Alternatively, it contains a chiral multiplet and an anti-chiral multiplet, of opposite $U(1)_t$ charge, but both transform in the fundamental representation. The above way of writing $\Phi_H(x)$ is adapted to the second viewpoint see section 5 for a related discussion. 

There is a universal contribution, for any $U(N)$ gauge group, coming from the vector multiplet:

\beq
\Phi_{V}(x) = \prod_{1\leq i <j \leq N} \prod_{n=0}^{\infty} { 1-q^n \; e^{x_i-x_j}\over 1-q^n\, t \;e^{x_i-x_j}}.
\eeq
The numerator is the W-boson contribution, and the denominator is the contribution from the adjoint of mass scalar of mass $\epsilon$.

The partition function takes the form of a matrix model, with the vector multiplet providing the measure factor

\beq\nonumber \Phi_V(x) = \Delta^2_{q,t}(x).
\eeq
For $q=t$, this coincides with the Haar measure on the unitary group $U(N)$. For general $q,t$, it does not come from fixing any gauge symmetry -- however, it does come from the $U(N)$ gauge symmetry of the physical $U(N)$ gauge theory in 3d.
With these definitions, the partition function looks like a matrix average of operators
\beq
{1\over N!} \int d^N x\; \Delta^2_{q,t}(x) \; \prod_{a=1}^{{ M}} \Phi_{H_a}(x)  \; e^{k \,{\rm Tr}x^2/2\hbar + \zeta \,{\rm Tr} x/\hbar} .
\eeq
The matrix model is of the refined Chern-Simons type \cite{AS, AS2}. The relationship is not coincidental: by taking limits, the physical theory studied in \cite{AS} can be recovered.
For the most part, we will focus on the $k=0$ case, as only the $k=0$ theory has a $q$-Liouville interpretation. For $q=t$, this becomes the Chern-Simons matrix model of \cite{Marino, Aganagic:2002wv}.

\subsection{The vacua, integration cycles and residues}

We need to determine the contour of integration to fully specify the path integral.
The choice of a contour in the matrix model corresponds to the choice of boundary conditions at infinity in the space where the gauge theory lives \cite{Cheng:2010yw}. At infinity, fields have to approach a vacuum of the theory. For small $q$ and $t$, the vacua are the critical points of $W(x)$.

In a theory with $M$ hypermultiplets, there are $M$ vacua (as we will see in section 5, for small mass splitting and $\zeta$, the vacua are located near $x\sim m_a^{\pm}$, so quite literally, there is a correspondence between the hypermultiplets and the vacua).
Splitting the $N$ eigenvalues so that $N_a$ of them approach the $a$-th critical point, we break the gauge group,

$$
U(N) \qquad \rightarrow \qquad U(N_1)\times \ldots \times U(N_M).
$$
We can think of all the quantities appearing in the potential as real; then the integration is along the real $x$ axis. To fully specify the contour of integration, we need to prescribe how we go around the poles in the integrand. The integral can be computed by residues, with slightly different prescriptions for how we go around the poles for the different gauge groups. In this way, we get $M$ distinct contours
${\cal C}_{N_1, \ldots, N_M}$, and with them the partition function,

$$
{\cal Z}_{{\cal G}_{3d}}({\vec N})={1\over \prod_{a=1}^M N_a!}\oint_{{\cal C}_{N_1, \ldots, N_M}} d^Nx \;\Delta_{q,t}(x)^2\;
\prod_{a=1}^{{ M}} \Phi_{H_a}(x)   \; e^{-\zeta \,{\rm Tr} x/\hbar}.
$$
Dividing by $N_a!$ corresponds to dividing by the residual gauge symmetry, in other words, to permuting the $N_a$ eigenvalues in each of the vacua.

We will see in section 6 that the residue integrals have a physical interpretation in fact. The 3d theory we are studying arises as a theory on charge ${\vec N}=(N_1, \ldots, N_M)$ vortices in the 5d ${\cal N}=1$ $U(M)$ gauge theory with $2M$ hypermultiplets \cite{DH1,DH2}. The partition function of the 3d theory is the integral over the moduli space of vortices in the 5d theory -- when we put the theory in the 5d $\Omega$-background. The integral localizes to fixed points in the vortex moduli space. When we express the partition function of the 3d gauge theory as a residues integral, we make its 5d origin manifest\footnote{The ${\cal G}_{3d}$ partition function is closely related to that in \cite{Shadchin:2006yz} which studied localization on the moduli space of charge $N$ vortices. The main difference is that \cite{Shadchin:2006yz} took the vortices to be zero dimensional, whereas for us they have dimension three.}.

\subsection{$M=2$ Example}

${\cal G}_{3d}$ is a $U(N)$ ${\cal N}=4$ gauge theory coupled to two flavors, $H_{1,2}$ in the fundamental representation. The mass deformations and the coupling break supersymmetry to ${\cal N}=2$.
The partition function equals

$$
Z_{{\cal G}_{3d}} = {1\over  N!}\int  d^Nx \;\Delta^2_{q,t}(x) \; \Phi_{H_1}(x)  \Phi_{{H}_2}(x)  \; e^{-\zeta {\rm Tr} x/\hbar}.
$$
Let us write the contribution of the hypermultiplets as

$$
\Phi_{H_1}(x)=  \prod_{n=0}^{\infty} { 1- q^n u \;  e^{x-m_{1,-}} \over 1-q^n  v\, e^{x-m_{1,+}}  } ,\qquad \Phi_{H_2}(x) =\prod_{n=0}^{\infty}
{ 1-  q^n u \; e^{x-m_{2,-}}\over 1-q^n v \;e^{x-{m}_{2,+} }}.
$$
For small mass splittings, there are two vacua near $x\sim m_{1, \pm}$ and $x\sim m_{2, \pm}$.
\subsubsection{The Matrix Integral }
Splitting the integral into two groups of eigenvalues, one related to the first, the other to the second critical point
breaks the symmetry of the theory as
$$
U(N)\qquad \rightarrow \qquad U(N_1)\times U(N_2).
$$
Now, let us define
$$
z_1=e^{- m_{1,+}} v , \;\;  \;\;
z_2= e^{-m_{2,+}}v  ,
$$
$$
\alpha_0=\zeta/\hbar-1  ,\qquad q^{\alpha_1}=e^{m_{1,+}-m_{1,-}} t, \qquad q^{\alpha_2}=e^{m_{2,+}-m_{2,-}}t  ,
$$
and change variables from $x$ to $y$, with
$$
y= e^{-x}.
$$
The partition function becomes the following:

\beq\label{3d4pt}
{\cal Z}_{{\cal G}_{3d}} (N_1,N_2)= {1\over N_1! N_2! }\oint_{{\cal C}_{1},{\cal C}_{2}} d^{N} y \; \Delta^2_{q,t}(y) \,   V_0(y) \; {V}_1(y)\; V_2(y), \;
\eeq
(up to an irrelevant  factor of $(-1)^N$ from the Jacobian).
In the new variables, the measure factors equal

\beq\label{meas}
\Delta^2_{q,t}(y) = \prod_{1\leq i\neq j \leq N} {\varphi(  y_i/y_j)  \over \varphi(t\,    y_i/y_j)},
\eeq
where $\varphi(y)$ is the quantum dilogaritm function:

$$
\varphi( y) =  \prod_{n=0}^{\infty} {( 1- q^n \,y)}.
$$
The potentials $V_a$ (for $a=0,1,2$) come from

$$
  \Phi_{H_1}(x)  \Phi_{{ H}_2}(x)  \; e^{-\zeta\, {\rm Tr} x/\hbar},
$$
together with the Jacobian associated to the change of variables:
$$
V_0(y) = \prod_{i=1}^N y_i^{\alpha_0},
$$
$$
{V}_{1} (y)=  \prod_{i=1}^{N}  {\varphi(q^{\alpha_1}  z_1/y_i)\over\varphi(z_1/y_i  )} ,\qquad
{V}_{2} (y)=  \prod_{i=1}^{N}  {\varphi(q^{\alpha_2}  z_2/y_i)\over\varphi(z_2/y_i  )}.
$$

As we will see in the following section, the partition function in \eqref{3d4pt} manifestly equals the 4-point conform block of $q$-Liouville, viewed as the Dotsenko-Fateev integral. This illustrates our claim that the Dotsenko-Fateev approach to the Liouville conformal blocks computes three dimensional ${\cal G}_{3d}$ gauge theory partition function on $M_q=({\mathbb C}\times S^1)_q.$

We will now spell out the choice of contours in evaluating the integral. We will see that, evaluating the contours by residues, the sum over poles rewrites the 3d partition function of ${\cal G}_{3d}$ as the Nekrasov partition function of the 5d gauge theory ${\cal G}_{5d}$.
\subsubsection{Summing over Residues}

The quantum dilogarithm  $\varphi( y) =  \prod_{n=0}^{\infty} {( 1- q^n \,y)}$ has zeros at $y=q^{-n}$, hence the integrands have poles.
The contour is chosen so as to pick up the residues of the poles of the integrand. For the first group of
$N_1$ eigenvalues we choose the contour that runs from $0$ to $z_1$, circling the poles at

$$ y=  q^n\,z_1, \qquad n=0, 1,\ldots .
$$
For the second group of $N_2$ eigenvalues, we will circle the poles at

$$\qquad y= q^{n}\, z_2, \qquad n=0,1,\ldots.
$$
instead. For $|t|, |q|<1$,  the poles interpolate between $y=0$ and $y= z_a$, and the contours ${\cal C}_a$ circle around the interval (this is also where the critical points of the integral are located). However, not all the poles contribute -- the numerator measure factor eliminates some. The poles that end up contributing are labeled by two Young diagrams, $P$ and $R$,
$$
(P,R)
$$
with at most $N_1$ and $N_2$ rows respectively:
\beq\label{yp}
(y_1, \ldots ,y_{N_1})_{P}=  (q^{P_1} t^{N_1-1} \,z_1 , q^{P_2} t^{N_1-2} \,z_1 ,\ldots  , q^{P_{N_1}} \,z_1 ),
\eeq
and
\beq\label{zp}
(y_{N_1+1}, \ldots , y_{N})_{R}=  (q^{R_1} t^{N_2-1}\,z_2, q^{R_2} t^{N_2-2} \,z_2,\ldots  , q^{R_{N_2}} \,z_2 ).
\eeq
This breaks the permutation symmetry of the two groups of eigenvalues, and soaks up the factorials.

To see that these are the poles contributing, focus first on the terms in the integrand corresponding to $n=0$. To fix the permutation symmetry we order $|y_i/y_j|\leq 1$ for $i\leq j$.  If $|t|<1$, there are poles within the contour
coming from the measure $\Delta^2_{q,t}(y)$  and the $\varphi(z_1/y)$ factor which occur at

$$y_i/y_{i+1} = t,   \qquad y_{N_1}=z_1.
$$
This gives
$y_i = z_1 t^{N_1-i}$, for $i$ running from $1$ to $N_1-1$.
This is not the only set of poles within the contour. Rather, there are other poles, in each of the $N_1$ $y$ variables,  shifted relative to the one we just found by $q^n$. Assuming $|q|\leq 1$, these correspond to

\beq\label{ppole}
y_i =  q^{P_i} t^{N_1 - i} z_1, \qquad i=1,\ldots, N_1
\eeq
for an arbitrary Young diagram $P$ with $N_1$ rows. Alternatively, we could have started with the poles originating from $\varphi(z_1/y)$ and noticed that the shift by $t^{N_1-i}$ is necessary for the pole to have non-zero residue; otherwise, the numerator of $\Delta_{q,t}^2$ vanishes.

An analogous consideration holds for the second group of $N_2$ eigenvalues. We get poles labeled by Young diagram $R$ with $N_2$ rows:

\beq\label{rpole}
y_{i+N_1} =  q^{R_i} t^{N_2 - i} z_2, \qquad i=1,\ldots, N_2.
\eeq
It is easy to see these are the only nontrivial contributions to the integral.

This replaces the integrals

$$
{1\over N_1! }\oint_{{\cal C}_{1}} d^{N_1} y  \;\times \;  {1\over N_2! } \oint_{{\cal C}_{2}} d^{N_2} y\qquad \rightarrow \qquad \sum_{P, R}
$$
by sums over the residues of the poles labeled by a pair of partitions $P, R$ at

$$
{\vec y} = {\vec y}_{(P,R)}, \qquad i=1, \ldots, N
$$
where ${\vec y}_{(P,R)}$ is defined in equations \eqref{ppole} and \eqref{rpole}.

To find the residues at the poles, we will take a shortcut. While we cannot simply evaluate the integrand at the pole, as we would get infinity, it is easy to show that the {\it ratio} of the values of the integrand at successive poles is finite. Thus, if we are interested in the value of the ratio of residues only, as opposed to the residues themselves, this ratio is the same as the ratio of the integrand at the two poles. Thus, the partition function can be written as

$$
Z_{{\cal G}_{3d}} =r \sum_{P,R} \; q^{\alpha_0(|R|+|P|)} \; V_{P,R},
$$
where $V_{P,R}$ is the residue at the pole labeled by $(P,R)$, normalized by the residue of the  $(\varnothing, \varnothing)$ pole; we denoted the value of the latter by $r$. $V_{PR}$ is computed by simply taking the ratio of the integrands, at the two poles:

$$
V_{P,R} = {[\Delta^2_{q,t}(y)\Psi_{H_1}(y)\Psi_{H_2}(y)]_{{\vec y} = {\vec y}_{(P,R)}}\;\over
[\Delta^2_{q,t}(y)\Psi_{H_1}(y)\Psi_{H_2}(y)]_{{\vec y} = {\vec y}_{(\varnothing, \varnothing)}}}.
$$

We will not need the actual value of $r$; to match to the 5d instanton partition function in section 5, we will normalize the partition function to set $r$ to $1$. This is not the full answer for the 5d partition function, of course; there are also the perturbative and the one loop factors. The product of the two should be accounted by the product of $r$, the residue at the $(\varnothing, \varnothing)$ pole, together with the contribution that is not captured by the gauge theory -- lets denote it by $C$. As discussed in section 6, the duality relating ${\cal G}_{3d}$ and ${\cal G}_{5d}$ is really a large $N$ duality in M-theory. The theory before the transition goes beyond the 3d gauge theory itself: what the gauge theory misses is the partition function of the closed string background without branes -- this is the $C$ factor. In general, $C$ is not trivial. This factor is also needed in the next section, when we relate ${\cal G}_{3d}$ to Liouville. We will find that the screening charge integrals of $q$-Liouville equal the 3d gauge theory partition function ${\cal Z}_{3d}$. But, the $q$-Liouville conformal block has a factor beyond the screening charge integrals. This factor is precisely $C$, the contribution of the closed string geometry to the partition function.

\subsection{$M+2$ point function}

Let us now consider ${\cal G}_{3d}$ with an arbitrary number $M$  of hypermultiplets. The partition function of the theory, with $U(N)$ gauge group is:

\beq\label{si}
{\cal Z}_{{\cal G}_{3d}}= {1\over N!} \int d^Nx \;\Delta_{q,t}(x)^2\;
\prod_{a=1}^{{ M}} \Phi_{H_a}(x)  \; e^{-\zeta \,{\rm Tr} x/\hbar} .
\eeq
where

\beq
\Phi_{H_a}(x)= \prod_{1\leq i \leq N}  \prod_{n=0}^{\infty}{ 1 -q^n u \; e^{x_i-m_{a,-}}\over 1-q^n v \; e^{x_i-m_{a,+}}}.
\eeq

\subsubsection{The Matrix Integral}
Letting as before $e^{-x}=y$,
the matrix integral becomes

\beq\label{liouville}
{\cal Z}_{{\cal G}_{3d}}({\vec N}) ={1\over \prod_{a=1}^M N_a! }\oint_{{\cal C}_{1}, \ldots {\cal C}_M} d^{N} y\; \Delta_{q,t}^2(y)\; \prod_{a=0}^M {V}_a(y),
\eeq
where
\beq\label{pot0}
V_0(y) = \prod_{i=1}^{N} y_i^{\alpha_0},
\eeq
and
\beq\label{pota}
{V}_{a} (y)=  \prod_{i=1}^{N}  {\varphi(q^{\alpha_a}  z_a/y_i)\over\varphi(z_a/y_i  )}.
\eeq
Similar to the $M=2$ case, we define
\beq\label{potp}
z_a=e^{- m_{a,+}} v , \qquad \alpha_0=\zeta/\hbar-1  ,\qquad q^{\alpha_a}=e^{m_{a,+}-m_{a,-}} t,
\eeq
for $a=1, \ldots, M$.

The theory has $M$ vacua, near $x\sim z_a$, which give rise to $M$ contours
$${\cal C}_1, \ldots, {\cal C}_M.
$$
For $|q|, |t|<1$, the contours run around the intervals in the complex $y=e^{-x}$ plane:  ${\cal C}_a$ circles the interval from $y=0$ to $y=z_a $, where $z_a$ is the location of a pole in the integral corresponding to a chiral multiplet going massless.

In the next section, we will see that the matrix integral \eqref{liouville} is the Dotsenko-Fateev integral for the $q$-Liouville conformal block on a sphere with $M+2$ punctures. Next, we will show how to compute the integral using residues. In section 4, we will show that it equals the Nekrasov partition function of the 5d ${\cal N}=1$ gauge theory ${\cal G}_{5d}$, the $U(M)$ gauge theory with $2M$ flavors.
\subsubsection{Sum over Residues}
Computing the integral by residues localizes it to points
\beq\label{amp}
{\vec y} = {\vec y}_{{\vec R}}.
\eeq
labeled by $M$-tuples of Young diagrams

\beq\label{bmp}
{\vec R} = (R_1, \ldots, R_a, \ldots, R_M),
\eeq
where $R_a$ has at most $N_a$ rows. The poles corresponding to the $a$-th group of variables are at

\beq\label{cmp}
y_{(N_1+\ldots +N_{a-1})+ i} = q^{R_{a,i}} t^{N_a - i}z_a,
\eeq
where $i$ runs from $1$ to $N_a$ and $a$ from $1$ to $M$.

The sum over the residues of the integral becomes the sum over the Young diagrams

$$
\prod_{a=1}^M {1\over N_a! }\;\oint_{{\cal C}_{1}, \ldots {\cal C}_M} d^{N} y\qquad \rightarrow \qquad \sum_{{\vec R}}.
$$
The partition function is the sum

$$
{\cal Z}_{{\cal G}_{3d}}({\vec N}, {\vec z}, {\vec \alpha} ) = r \sum_{{\vec R}}\; I^{3d}_{\vec R}
$$
where

\beq\label{3dsummand}
I^{3d}_{\vec R} =q^{\alpha_0|{\vec R}|} \;V_{{\vec R}}({\vec z}, {\vec \alpha}).
\eeq
Here, $V_{{\vec R}}$ is the residue at ${\vec R}$, normalized by the residue at ${\vec \varnothing}$; we denoted the latter by $r$. As in the $M=2$ case, the normalized residue is easier to compute than its actual value: it is obtained by evaluating the integrand of the matrix model at the two poles; the latter is finite

\beq\label{3dsummand}
V_{{\vec R}} = {[\Delta^2_{q,t}(y)\;
\prod_{a=0}^{{ M}} V_a(y)]_{{\vec y} = {\vec y}_{\vec R}}\; \over \;[\Delta^2_{q,t}(y)\;
\prod_{a=0}^{{ M}} V_a(y)]_{{\vec y} = {\vec y}_{\vec \varnothing} }}.
\eeq
In addition,

\beq
q^{\alpha_0|{\vec R}|}
={[V_0(y)]_{{\vec y} = {\vec y}_{\vec R}}\over [V_0(y)]_{{\vec y} = {\vec y}_{\vec \varnothing} }}.
\eeq
Finally, recall that the origin of the measure factor  $\Delta^2_{q,t}(y)$ is the vector multiplet contribution to the partition function,

\beq
 \Delta^2_{q,t}(y)=\Phi_V(y),
\eeq
and the potentials $V_a(y)$, $a=1,\ldots, M$ are the hypermultiplet contributions,

$$
\prod_{a=1}^{{ M}} V_a(y)=\prod_{a=1}^M \Phi_{H_a}(y).
$$
We will find this structure replicated in the 5d gauge theory ${\cal G}_{5d}$.

%
%
%

\section{Liouville and its $q$-deformation}

In the previous section, we considered the 3d gauge theory ${\cal G}_{3d}$, with $M$ flavors in the fundamental representation. In this section, we will show that the partition function of the theory on $M_q = ({\mathbb C} \times S^1)_q$ is the Dotsenko-Fateev matrix model for an $M+2$ conformal block of $q$-Liouville on a sphere with $M+2$ punctures.

The conformal blocks of Liouville field theory possess several complementary representations that highlight various properties. For example, the operator product expansion procedure provides a \emph{series} representation for the conformal blocks in terms of sums over the Virasoro descendants of the internal primary fields, parametrized by tuples of Young diagrams (see \cite{Marshakov:2009gs} for a recent exposition). A representation that plays a crucial role in our work is an \emph{integral} representation in terms of free fields, discovered by Dotsenko and Fateev \cite{Dotsenko:1984nm}, and recently revisited in the papers \cite{Dijkgraaf:2009pc, Schiappa:2009cc, Mironov:2009ib, Mironov:2010zs, Mironov:2010su, Mironov:2010pi, Awata:2010yy, Zhang:2011au} in the context of understanding the AGT conjecture. Consider the conformal block

$$
{\cal B}_{{\alpha_0}, \ldots, \alpha_{M+1}}(z_1, \ldots z_{M}),
$$
corresponding to a
sphere with $M+2$ Liouville primaries of momenta
$$\alpha_0, \ldots, \alpha_{M+1}$$ inserted at
$z=z_0, \ldots, z_{M+1}.$
By conformal invariance, we can set $z_0=0$, and $z_1= 1$, and $z_{M+1}=\infty$; it is somewhat more convenient to leave $z_1$ arbitrary  for now, so that the insertion points are at
$$
0, z_1, \ldots, z_M, \infty.
$$

Liouville field theory has the Coulomb gas representation
$$
{\cal B}_{{\alpha_0}, \ldots, \alpha_{M+1}}(z_1, \ldots z_{M}) =\oint dy_1\ldots \oint dy_N\; \langle V_{\alpha_0} (0)\ldots V_{\alpha_M}(z_M)\;
S(y_1)\ldots S (y_N)\rangle,
$$
where insertion of a primary with momentum $\alpha$
at $z$ is realized by the vertex operator
$$
V_{\alpha}(z) = : e^{\alpha \phi(z)/\beta}:,
$$
in the presence of $N$ screening charge integrals
$$
\oint dy \;S (y) = \oint dy\, : e^{2 \phi(y)}:
$$
Here $\phi(z)$ is a free field,
$$
 \phi(z) =  {\hat q} + {\hat p} \log z + \sum\limits_{n \neq 0} \dfrac{1}{n} {\hat J}_n z^n,
$$
with canonical commutation relations
\begin{align*}
[{\hat J}_n, {\hat J}_m] = {\beta\over 2}  \;n \delta_{n+m,0}, \ \ \ [ {\hat p}, {\hat q} ] = {\beta\over 2}.
\end{align*}

\begin{figure}[h]
  \begin{center}
    \includegraphics[width=0.6\textwidth]{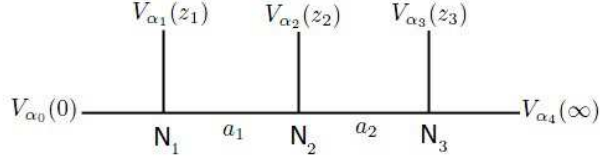}
  \end{center}
  \caption{The ``comb" diagram and parameters of the associated block.}\label{comb}
\end{figure}

The $N$ screening charges play the role of the insertion at infinity and fix the momentum $\alpha_{M+1}$ by
$$
\sum_{a=0}^{M+1} \alpha_a + 2 \beta N = 2-2\beta.
$$
To fully determine the conformal block we have to specify the intermediate momenta. The choice of the intermediate momenta corresponds to the choice of contours for the screening charges. As explained in \cite{Mironov:2010zs}, one splits the $N$ screening integrals into $M$ groups with $N_a$ screening integrals each:
$$
N\;\;  = \;\; \sum_{a=1}^M \;\;N_a,
$$
with the contour for the $a$-th group encircling the segments ${\cal C}_{a}$
$${\cal C}_{a}:\qquad  [0, z_a], \qquad a=1, \ldots, M,
$$
where $z$'s  are the positions of the $M$ punctures on the sphere away from $0$, $\infty$. This sets the corresponding internal momenta $a_b$ to be given by
$$a_1= \alpha_0 + \alpha_1 + 2 \beta N_1, \qquad a_b = a_{b-1} + \alpha_{b} + 2 \beta N_b,\qquad b=2, \ldots, M-1.
$$
This is in accordance with the main principle of free field integrals: the amount of charge conservation violation $a_{b} - a_{b-1} - \alpha_{b}$ in each trivalent vertex of the diagram is equal to the total charge $2 \beta N_b$ of screenings associated with this vertex. The particular conformal block that one obtains by this Dotsenko-Fateev integral is of the comb type, as on the Figure \ref{comb}.

The evaluation of the correlator in the r.h.s. is straightforward: using the Wick theorem or any of its analogues, one finds
\begin{align*}
\nonumber \langle V_{\alpha}(z) V_{\alpha^{\prime}}(z^{\prime}) \rangle = (z - z^{\prime})^{\frac{\alpha \alpha^{\prime}}{2 \beta}}, \\
\nonumber \langle V_{\alpha}(z) S(z^{\prime}) \rangle = (z - z^{\prime})^{\alpha}, \\
\langle S(z) S(z^{\prime}) \rangle = (z - z^{\prime})^{2\beta},
\end{align*}
and obtains the following Dotsenko-Fateev integral\footnote{The
equality holds up to a proportionality factor
$C = \prod\limits_{1\leq a<b\leq M} \left(z_a-z_b\right)^{\frac{\alpha_a \alpha_b}{2 \beta}}.$}
\begin{align*}
{\cal B}_{{\alpha_0}, \ldots, \alpha_{M+1}}(z_1, \ldots z_{M})= {C\over \prod_{a=1}^M N_a!}\oint_{{\vec{\gamma}}} d^Ny\; \Delta_{\beta}^{2}(y) \;\prod_{a=0}^M V_a(y),
\end{align*}
where the measure is the $\beta$-deformed Vandermonde,

$$
\Delta_{\beta}^{2}(y) = \prod_{1\leq i<j\leq N}(y_i-y_j)^{2\beta},
$$
and the potential equals
$$
V_a(y) =  \prod\limits_{i=1}^{N} (y_i - z_a)^{\alpha_a}.
$$

\subsection{$q$-deformed Liouville}
Generalization to $q$-deformed Liouville theory involves the deformation of the basic ingredients: the vertex and screening operators. This deformation has been worked out in the 90's in \cite{Shiraishi:1995rp, Awata:1996xt} using the quantum group techniques:

\begin{align*}
& V_{\alpha}(z) = \ : \exp\left( \frac{\alpha}{\beta} {\hat q} +\frac{\alpha}{\beta} {\hat p} \log z + \sum\limits_{n \neq 0} \dfrac{1 - q^{\alpha n}}{1 - t^n} \dfrac{1}{n} {\hat J}_n z^n \right) : \\
& S(z) = \ : \exp\left( 2 {\hat q} + 2 {\hat p} \log z + \sum\limits_{n \neq 0} \dfrac{1 + (q/t)^n}{n} {\hat J}_n z^n \right) :
\end{align*}
The previous commutation relations between the modes are now

\begin{align*}
[{\hat J}_n, {\hat J}_m] = \dfrac{1}{1 + (q/t)^n} \frac{1 - t^n}{1 - q^n} \ n\,  \delta_{n+m,0}, \ \ \ [ {\hat p}, {\hat q} ] = \frac{\beta}{2}.
\end{align*}
Just as before, using these commutation relations, one computes the correlator and obtains the following Dotsenko-Fateev integral:

\begin{align}\label{lcv}
{\cal B}_{\alpha_0,\alpha_1\ldots,\alpha_{M+1}}(z_1, \ldots z_{M}) =  {{C} \over \prod_{a=1}^M  N_a!} \oint_{{\cal C}_1, \ldots, {\cal C}_M} d^{N}y\; \Delta^2_{q,t}(y) \; \prod_{a=0}^M V_a(y; z_a)
\end{align}
where the measure is the $q,t$-deformed Vandermonde

$$
\Delta_{q,t}^2(y) = \prod_{1\leq i\neq j\leq N} {\varphi(y_i/y_j)\over \varphi(t\; y_i /y_j)},
$$
and the potential equals
$$
V_a(y; z_a) =  \prod\limits_{i=1}^{N} \dfrac{ \varphi\big(q^{\alpha_a} {z_a/y_i}\big) }{ \varphi\big(z_a/y_i\big) }.
$$
In particular, using the properties of the quantum dilogarithm, it is easy to find that $V_0( y, 0) = (y_1 \ldots y_N)^{\alpha_0}$. As in the undeformed case, the relation holds up to a constant of proportionality $C$. In this paper, we avoid detailed consideration of this normalization constant. The meaning of the constant $C$, on the Liouville side, is to account for all possible two-point functions between the vertex operators $V_{\alpha}(z_a)$. In the 5d gauge theory language, this means that we limit ourselves only to the instanton part of the 5d gauge theory partition function. On the 3d gauge theory side, this is the contribution of the bulk theory, not captured by the theory on the vortex, as we discussed in section 2.

The matrix integral \eqref{lcv} manifestly equals the partition function of ${\cal G}_{3d}$, as written in \eqref{liouville}, with potentials defined in
\eqref{pot0}, \eqref{pota}, \eqref{potp} and the measure factor from \eqref{meas}. Like in the undeformed case, the $N$ eigenvalues are grouped into sets of size $N_a$, $a=1,\ldots, M$, by the choice of contours they get integrated over. The contours of integration \emph{are the same as} in the undeformed case -- encircling the segments $[0, z_a]$. The $q$ deformation affects the operators and the algebra, but not the contours.  It is important to emphasize that these contours agree with the alternative approach \cite{Mironov:2011dk} where the Dotsenko-Fateev integrals are replaced by Jackson $q$-integrals: in our picture, the latter are the residue sums for the former.

\section{The 5d Gauge Theory ${\cal G}_{5d}$}

In section $2$, we studied the 3d theory ${\cal G}_{3d}$ with $M$ hypermultiplets in the fundamental representation.  The partition function of the theory ${\cal Z}_{{\cal G}_{3d}}$ in 3d $\Omega$-background localizes to a sum over residues, labeled by $M$-tuples of Young diagrams. In this section, we will show that residue sum in ${\cal Z}_{{\cal G}_{3d}}$ is equal to the instanton sum of a 5d theory ${\cal G}_{5d}$ in $\Omega$-background, studied in \cite{Moore:1997dj, Losev:1997wp, Nekrasov:2002qd, Nekrasov:2003rj}, at integer values of Coulomb branch parameters. The 5d gauge theory ${\cal G}_{5d}$ has  ${\cal N}=1$ supersymmetry,  a $U(M)$ gauge group with $M$ hypermultiplets in the fundamental representation, and $M$ in the anti-fundamental. The 5d $\Omega$-background is
$$M_{q,t} = ({\mathbb C}\times {\mathbb C}\times S^1)_{q,t}.
$$
The subscript denotes that,  as one goes around the $S^1$, one rotates the two complex planes by $q$ and $t^{-1}.$  These are paired together with the 5d $U(1)_R$ symmetry twist, to preserve supersymmetry. The 5d gauge theory partition function in this background is the trace

\beq\label{5dtrace}
{\cal Z}_{{\cal G}_{5d}}={\rm Tr} (-1)^F q^{S_1-S_R} t^{S_R-S_2},
\eeq
where the trace is computed in the quantum mechanics on the moduli space of instantons in ${\cal G}_{5d}$ (see \cite{Nekrasov:2004vw} for a review), $S_{1,2}$ are the generators of rotations of the two complex planes, and $S_R$ is the $U(1)_R$ symmetry generator. The fact that the 3d and the 5d traces look so similar is not a coincidence, as we will explain in section 6.

The instanton partition function of ${\cal G}_{5d}$ in this background is  \cite{Moore:1997dj, Losev:1997wp, Nekrasov:2002qd, Nekrasov:2003rj,Nekrasov:2004vw}, takes the form of a sum 

\beq\label{bN}
{\cal Z}_{{\cal G}_{5d}} = \sum_{\vec R} I^{5d}_{\vec R},
\eeq
over $M$-touples of partitions

$$
{\vec R} = (R_1, \ldots , R_M),
$$ 
labeling fixed points in the instanton moduli space, where a given fixed point contributes 

\beq\label{5dI}
I^{5d}_{\vec R}= \;q^{\xi \cdot \vec R} \;\; T^k_{{\vec R}} \;\;Z_{{\vec R}}({\vec e}, \,{\vec f}).
\eeq
In the partition function, $Z_{{\vec R}}$ depends on $M$ Coulomb branch moduli, encoded in an $M$-vector ${\vec e}$, and the $2M$ parameters ${\vec f}$, which are the masses of the $M$ fundamental hypermultiplets $H$ and the $M$ anti-fundamental hypermultiplets $H^{\dagger}$.  The sum is over $M$-tuples of Young diagrams, $(R_1, \ldots, R_M)$ labeling the fixed points in the instanton moduli space. The instanton charge is the net number of boxes $|\vec R|$ in the $R$'s. The instanton counting parameter, related to the gauge coupling of the theory, is $q^{\xi}$. The parameter $k$ is the 5d Chern-Simons level; since we have $q$-Liouville application in mind, we will set it to zero in what follows.

The contribution

$$Z_{\vec R} =z_{V, {\vec R}} \times z_{H, {\vec R}} \times z_{H^{\dagger}, {\vec R}},
$$
of each fixed point is a product over the contributions of  the $U(M)$ vector multiplets,  the fundamental hypermultiplets $H$, and the anti-fundamental hypermultiplets $H^{\dagger}$ in ${\cal G}_{5d}$.
The vector multiplet contributes

$$
z_{V, {\vec R}}= \prod_{1\leq a,b\leq M}[N_{R_a R_b}(e_a/e_b)]^{-1}.
$$
The $M$ fundamental hypermultiplets contribute

$$
z_{H, {\vec R}} = \prod_{1\leq a \leq M} \prod_{1\leq b \leq M}N_{\varnothing R_b}( v f_{a}/e_b),
$$
and the $M$ anti-fundamentals give
$$
z_{H^{\dagger}, {\vec R}} = \prod_{1\leq a \leq M} \prod_{1\leq b \leq M}N_{R_a \varnothing }( v e_a/{f_{b+M}}).
$$
The basic building block is the Nekrasov function
\begin{align*}
N_{RP}(Q) = \prod\limits_{i = 1}^{\infty} \prod\limits_{j = 1}^{\infty}
\dfrac{\varphi\big( Q q^{R_i-P_j} t^{j - i + 1} \big)}{\varphi\big( Q q^{R_i-P_j} t^{j - i} \big)} \
\dfrac{\varphi\big( Q t^{j - i} \big)}{\varphi\big( Q t^{j - i + 1} \big)}.
\end{align*}
with $\varphi(x) = \prod\limits_{n=0}^{\infty}(1-q^n x)$ being the quantum dilogarithm we previously introduced. Furthermore,
$ T_R =(-1)^{|R|} q^{\Arrowvert R\Arrowvert/2}t^{-\Arrowvert R^t\Arrowvert/2}$, and $v = {(q/t)^{1/2}}$ as before (we use the conventions of \cite{Awata:2008ed}).

In what follows, it is good to keep in mind that there is no essential distinction between the fundamental and anti-fundamental hypermultiplets. By varying the Coulomb branch and the mass parameters, the real mass $m$ of the 5d hypermultiplet can go through zero. This exchanges the fundamental hypermultiplet of mass $m$ for an anti-fundamental of mass $-m$, while at the same time the 5d Chern-Simons level jumps by $1$ \cite{Witten5dphases}. This is reflected in the following relation between the anti-fundamental and the fundamental hypermultiplet contributions to the partition function:
\beq\label{faf}
z_{{H}^{\dagger}, {\vec R}}({\vec e}, m) =  z_{H, {\vec R}}({\vec e}, m) \;\times {T_R},
\eeq
where the equality holds up to a shift of $\vec{\xi}$, where
\beq\label{5dflip}\begin{aligned}
 z_{H, {\vec R}}&({\vec e}, m) =
\prod_{1\leq a\leq M} N_{\varnothing R_a}( v m/e_a),\\
  z_{H^{\dagger}, {\vec R}}&({\vec e}, m) =
\prod_{1\leq a \leq M} N_{R_a \varnothing }( v e_a/m),
\end{aligned}
\eeq
and the relation follows from the following identity
\begin{align}\label{iden}
N_{R_aR_b}(Q^{-1}) = N_{R_bR_a}(v^2 Q)\; {T_{R_a}\over{T}_{R_b}}\; ( Qv^{-1})^{|R_a|+|R_b|},
\end{align}
which the Nekrasov function satisfies. 
Which way of thinking about the theory is more natural depends on whether $|m/e_a|>1$ or $|m/e_a|<1$ . The partition function is analytic in its parameters, so all the different expressions are the same. In keeping with this, it is natural to think of all the $2M$ matter multiplets at the same footing,
and write the partition function, say, in terms of the fundamentals alone, whose masses run over $2M$ values, $f_a, f_{M+a}$, with $a=1, \ldots, M$.

In what follows, we will see that upon tuning the $M$ Coulomb branch parameters ${\vec e}$  to equal,  up to integer shifts, the mass parameters of $M$ of the hypermultiplets,
the instanton sum of ${\cal G}_{5d}$ and 
the residue sum of the ${\cal G}_{3d}$ partition function, defined in \eqref{3dsummand},

$$
I^{5d}_{\vec R}
=I^{3d}_{\vec R},
$$
coincide.

\subsection{Truncation at Integer Values}

We need two properties of Nekrasov functions. The first is that \cite{Awata:2008ed, DHG}

$$
N_{R\varnothing}(t^N) , \qquad N_{\varnothing R}( v^2 t^{-N})
$$
vanish if the Young diagram of $R$ has more that $N$ rows. Suppose the representation $R$ has no more than $N$ rows. Then, one can write the Nekrasov function as

\beq\label{noflip}
N_{R \varnothing}( Q ) = \prod\limits_{i = 1}^{N} \dfrac{\varphi(Q t^{1 - i})}{\varphi(Q q^{R_i} t^{1 - i})}.
\eeq
If we take $Q$ to be an integer power of $t$, $\varphi$'s in the numerator vanish for representations that get too big. We tune the $M$ Coulomb branch parameters to equal the masses of  $M$ of the fundamentals, up to an integer. Up to permutations, we can set these to be the ``first" $M$ fundamentals ordered so that

\beq\label{Coulomb}
e_a/ f_a = t^{N_a}/ v, \qquad a=1, \ldots, M.
\eeq
This parameter also measures the size of the cut in the Riemann surface; the latter closes up at $e_a/f_a=1$.
Given \eqref{Coulomb}, the index vanishes unless the corresponding Young diagram $R_a$ has at most $N_a$ rows:

$$
R_{a,i} = 0, \qquad i\geq N_a,
$$
since  $N_{\varnothing R_a}(v f_a/e_a)$ vanishes otherwise.

Secondly, if $R_a$ and $R_b$ have at most $N_a$, $N_b$ rows, respectively, then

$$
\begin{aligned}\label{trunc1}
N_{R_aR_b}(Q) &=
\prod\limits_{i = 1}^{N_a} \prod\limits_{j = 1}^{N_b}
\dfrac{\varphi\big(t\, Q\, q^{R_{a,i}-R_{b,j}} t^{j - i} \big)}
{\varphi\big(\;Q \, q^{R_{a,i}-R_{b,j}} t^{j - i}  \big)} \ \dfrac{\varphi\big(\; Q\, t^{j - i} \big)}{\varphi\big(t\,  Q\, t^{j - i} \big)}
\\
& \times
\left[N_{R_a\varnothing}\Big( t^{N_b}Q  \Big) N_{\varnothing R_b}\Big( t^{- N_a}Q \Big) \right]
\end{aligned}
$$
Furthermore, given \eqref{Coulomb}, the vector multiplet contribution

$$
z_{V, \vec R}({\vec e})= \prod_{1\leq a,b \leq M} N_{R_a R_b}(e_a/e_b),
$$
takes the form

\beq\begin{aligned}\label{3d5d}
z_{V,\vec R}({\vec e})&= \prod_{1\leq a,b \leq M} \prod\limits_{i \neq j} \dfrac{\varphi\big( \,(y_{\vec R})_{a,i} / (y_{\vec R})_{b,j} \, \big)}
{\varphi\big( \, t \, (y_{\vec R})_{a,i} / (y_{\vec R})_{b,j} \, \big) } \
\dfrac{\varphi\big( \, t \, (y_{\vec \varnothing})_{a,i} / (y_{\vec \varnothing})_{b,j} \, \big)}
{\varphi\big( \, (y_{\vec \varnothing})_{a,i} / (y_{\vec \varnothing})_{b,j} \, \big) } \\
& \times \prod_{1\leq a,b\leq M}
\left[N_{R_a\varnothing}\Big( t^{N_a}f_a/f_b  \Big) N_{\varnothing R_b}\Big( f_a/t^{N_b}f_b \Big) \right]^{-1},
\end{aligned}
\eeq
where the variables ${\vec y}_{\vec R}$:

$$
(y_{\vec R})_{a,i} = q^{R_{a,i}} t^{N_a - i} z_a,\qquad {a=1, \ldots M; \ i=1,\ldots N_a}
$$
were defined in equations \eqref{amp}, \eqref{bmp}, \eqref{cmp}, in the context of studying the ${\cal G}_{3d}$ partition function. Already at this level, we see the correspondence between the 3d and the 5d gauge theory emerge.  Note that the first line of \eqref{3d5d} is the 3d vector multiplet contribution to the 3d partition function, from the residue of the pole at ${\vec y} = {\vec y}_{\vec R}$; namely,

$$
{[\Delta^2_{q,t}(y)]_{{\vec y}= {\vec y}_{\vec R}}\over [\Delta^2_{q,t}(y)]_{{\vec y}= {\vec y}_{\vec \varnothing}}}
=
\prod_{1\leq a,b \leq M} \prod\limits_{i \neq j} \dfrac{\varphi\big( \, (y_{\vec R})_{a,i} / (y_{\vec R})_{b,j} \, \big)}
{\varphi\big( \, t \, (y_{\vec R})_{a,i} / (y_{\vec R})_{b,j} \, \big) } \,
\dfrac{\varphi\big( \, t \, (y_{\vec \varnothing})_{a,i} / (y_{\vec \varnothing})_{b,j} \, \big)}
{\varphi\big( \, (y_{\vec \varnothing})_{a,i} / (y_{\vec \varnothing})_{b,j} \, \big) }
$$
The relation holds provided $z_a/z_b = f_a/f_b$; we will see in a moment that in fact simply $z_a$ and $f_a$ equal, $z_a = f_a$.

\subsection{Emergence of the Matrix Integral}

We will now show that all the remaining terms combine to give the residue of the 3d gauge theory potential, defined in equations \eqref{pota}, \eqref{potp}, at ${\vec y}= {\vec y}_{\vec R},$

$$
\left[\prod_{a=1}^M V_a(y)\right]_{{\vec y}= {\vec y}_{\vec R}}.
$$
The terms
in the second line of \eqref{3d5d}

\beq\label{part1}
\prod_{1\leq a,b\leq M}
\left[N_{R_a\varnothing}\Big( t^{N_a}f_a/f_b  \Big) N_{\varnothing R_b}\Big( f_a/t^{N_b}f_b \Big)\right ]^{-1},
\eeq
naturally combine with the 5d matter multiplet contributions:

\beq\begin{aligned}\label{part2b}
 z_{H,{\vec R}}\times z_{H^{\dagger}, {\vec R}} &= \prod_{1\leq b \leq M}
\prod_{1\leq a\leq M}  N_{\varnothing R_a}( v f_{b+M}/e_a) f_{R_a} \times N_{\varnothing R_a}( v {f_{b}}/e_a).
\end{aligned}
\eeq
Using  $e_a/ f_{a} = t^{N_a}/ v$, we can rewrite the first term in \eqref{part2b} as

$$N_{\varnothing R_a}( v f_{b}/e_a)=N_{\varnothing R_a}( v^2 f_b/t^{N_a}f_a ) =N_{ R_a\varnothing }(  t^{N_a}f_a /f_b) (f_{R_a})^{-1},
$$
which makes it manifest that it cancels with the first term in \eqref{part1}.
Thus, taking the product of \eqref{part1} and \eqref{part2b}, we find

\beq\begin{aligned}\label{almost}
\prod_{1\leq a,b\leq M}
&{N_{\varnothing R_a}\Big( v f_b/e_a\Big)  N_{\varnothing R_a}\Big( v f_{b+M}/e_a\Big)\over N_{R_a\varnothing}\Big( t^{N_a}f_a/f_b  \Big) N_{\varnothing R_b}\Big( f_a/t^{N_b}f_b \Big) }\\ =
\prod_{1\leq a,b\leq M}
&{ N_{\varnothing R_a}\Big( v^2 f_{b+M}/t^{N_a} f_a \Big )\over  N_{\varnothing R_a}\Big( f_{b}/t^{N_a}f_a \Big) } \end{aligned}
\eeq
Now, if one recalls that
\beq\label{flip}
N_{\varnothing R}( Q ) = \prod\limits_{j = 1}^{N} \dfrac{\varphi(Q \, /\, q^{R_j} t^{-j})}{\varphi(Q\, /\,t^{-j})}.
\eeq
We can write \eqref{almost} as
\begin{align*}
{[\prod_{a=1}^M V_a(y)]_{{\vec y}={\vec y}_{\vec R}}\over [\prod_{a=1}^M V_a(y)]_{{\vec y}= {\vec y}_{\vec \varnothing}}}=
\prod_{1\leq a,b\leq M}  \prod\limits_{j = 1}^{N_a} \dfrac{\varphi(v^2  \,f_{M+b}/   \, f_a \, q^{R_{a,j}} t^{N_a - j})}{\varphi(\,f_b/\,f_a\, q^{R_{a,j}} t^{{N_a - j}})}
\dfrac{\varphi( \, f_b/\,f_a \,  t^{N_a - j})}{\varphi(v^2\, f_{M+b}/ \, f_a \,t^{N_a - j})},
\end{align*}
where

$$
V_a(y)=\prod_{i=1}^N{\varphi(q^{\alpha_a}z_a/ y_i)\over \varphi(z_a/ y_i)}, \qquad a=1, \ldots, M
$$
and we defined

\beq\label{parmap}
f_a  = z_a, \qquad f_{M+a} = q^{\alpha_a} z_a /v^{2}   \qquad a=1, \ldots, M.
\eeq

Putting together the vector and the hypermultiplet contributions, and letting 
$$
q^{\xi} = q^{\alpha_0}q, \qquad 
V_0(y)=\prod_{i=1}^Ny_i^{\alpha_0}.
$$ 
we find that the Nekrasov summand, defined in \eqref{bN}
 
$$
I^{5d}_{\vec R}= q^{\xi \cdot \vec R}  \;\;Z_{{\vec R}}({\vec e}, \,{\vec f}) ,
$$
equals

\beq\label{5d3d}
I^{5d}_{\vec R}= 
{[\Delta^2_{q,t}(y) \prod_{a=0}^M V_a(y)]_{{\vec y}={\vec y}_{\vec R}}\over [\Delta^2_{q,t}(y) \prod_{a=0}^M V_a(y)]_{{\vec y}= {\vec y}_{\vec \varnothing}}} =I^{3d}_{\vec R}
\eeq
the residue summand of the 3d partition function, defined in \eqref{3dsummand}.

\subsection{Summary: A triality}

To summarize the result of the last 3 sections: we have proven the equivalence of the partition functions
\beq\label{Triality}
{\cal Z}_{{\cal G}_{3d}}= {\cal B} =  {\cal Z}_{{\cal G}_{5d}},
\eeq
of the 3d gauge theory, ${\cal G}_{3d}$ with $M$ hypermultiplets, the $q$-Liouville conformal block ${\cal B}$ on a sphere with $M+2$ punctures, and the 5d $U(M)$ gauge theory ${\cal G}_{5d}$ instanton partition function ${\cal Z}_{{\cal G}_{5d}}$, the along the locus of the ${\cal G}_{5d}$ moduli space where the Coulomb-branch parameters of the 5d theory are set to equal the masses of $M$ of the $2M$ hypermultiplets, up to integer shifts

$$e_a =  t^{N_a}\,f_{a}\, /v, \qquad a=1,\ldots, M.
$$
The latter are set by the unbroken 3d gauge group, and the choice of screening charge integrals in Louville. The map of the remaining variables, between the 5d gauge theory, Liouville, and the 3d theory, is:

$$
f_{a}  = z_a = e^{-m_{+,a}} v, \qquad f_{a+M} = q^{\alpha_a} z_a /v^{2} = e^{-m_{-,a}}u/v^2
$$
for $a$ running from $1$ to $M$ and
$$
q^{\xi} = q^{\alpha_0}q=q^{\zeta/\hbar}.
$$

The first equality in \eqref{Triality} is manifest if we write ${\cal Z}_{{\cal G}_{3d}}$ as the integral over the Coulomb branch moduli of ${\cal G}_{3d}$,
%
The second is manifest if we compute ${\cal Z}_{{\cal G}_{3d}}$ by residues.
%
As we discuss in section 6, the 3d residue sum is a sum over vortices.

\section{Curves and Vertex Operators}

In the CFT/gauge theory correspondence of \cite{Gaiotto:2009we, AGT}, two Riemann surfaces play a role. One is the curve $C$, where the Liouville theory lives. The other is the Seiberg-Witten curve $\Sigma$ of the corresponding 4d gauge theory. $\Sigma$ is a double cover of $C$. In the $q$-Liouville case, the relevant double cover is the Seiberg-Witten curve of the 5d gauge theory.

In the present work, we have a triality between the 3d gauge theory ${\cal G}_{3d}$, $q$-Liouville, and the 5d gauge theory ${\cal G}_{5d}$. In this section, we will explain the geometry behind the triplet of correspondences.
 \subsection{$q$-Liouville and $C$}
The conformal blocks on an $M+2$ punctured sphere $C$ provide a pants decomposition of $C$. When we decompose $C$ into $M$ pants
$$
V_1, \;V_2, \ldots,V_M,
$$
the $M$ vertices one obtains are naturally associated with the chiral vertex operators \cite{MooreSeiberg}
$$V_a(z_a, \alpha_a),
$$
that arise in computing the Liouville conformal block. Here, $z_a$ labels the insertion point, and
$\alpha_a$ is the momentum of the chiral primary. The 3-point chiral block is in addition labeled by momentum flowing through. In our case, these momenta are quantized and labeled by the $M$ integers $N_a$.

\subsection{3d Gauge Theory and $\Sigma_+ \cup \Sigma_-$}
In terms of the 3d gauge theory ${\cal G}_{3d}$, the curve $C$ can be identified with the Coulomb branch of ${\cal G}_{3d}$. More precisely, for a $U(N)$ gauge theory, the Coulomb branch is classically an $N$-fold symmetric product, corresponding to choosing $N$ points on $C$. Insertion of a chiral vertex operator of Liouville

$$
V_a(z_a, \alpha_a) \qquad \longleftrightarrow \qquad \Phi_{H_a}(m_{a,+}, m_{a,-}).
$$
corresponds to coupling the 3d theory to a hypermultiplet $H_a$, with $z_a$ and $\alpha_a$ capturing the mass of $H_a$ and the mass splitting, respectively,

\beq\label{punc}
z_a=v \, e^{-m_a^+}, \qquad
q^{{\alpha}_a} = t \, e^{m_a^+ - m_a^-}.
\eeq

While the classical Coulomb branch of ${\cal G}_{3d}$ is described by $C$, the quantum moduli space of the theory is captured by a pair of Riemann surfaces $\Sigma_{+}\cup \Sigma_-$
which give a reducible, 2-component, double cover of ${C}$,
$$
\Sigma_{+}\cup \Sigma_- \qquad \rightarrow \qquad C.
$$
As we will see, $\Sigma_+$ and $\Sigma_-$ separately have no meaning; only their union does. The Riemann surfaces capture the effective twisted
superpotential of ${\cal G}_{3d}$, in the limit of vanishing $\epsilon$.
In this limit, only the hypermultiplets contribute and $W$ becomes the single trace $\sum_{i=1}^N W(x_i)$, determined by a function of one variable, which we will call $W(x)$. The superpotential\footnote{In the NS limit, \cite{NS, NS1} with vanishing $\hbar$ but non-vanishing $\epsilon$, the superpotential is more complicated since the adjoint and the W-bosons also contribute; see \cite{NS2, NRSnew} and \cite{DH1, DH2}.} $W(x)/\hbar$ is also the potential of the matrix model \eqref{basic}, in the limit $q=e^{\hbar} \rightarrow 1$. We split the ${\cal N}=2$ chiral multiplets contributing to ${  W}$ into two groups. We assign one group to $W_{+}(x)$, the other to $-W_-(x)$, in such a way that:

$$
{ W}(x)  = W_+(x)-W_-(x).
$$
The splitting is ambiguous of course, and it will lead to an ambiguity in defining $\Sigma_+$ and $\Sigma_-$ separately. Given $W_+(x)$ and $W_-(x)$, the two curves $\Sigma_{\pm}$ are defined as the set of points $p$ and $x$, which are related by

$$
\Sigma_{\pm} :\qquad e^{p} = e^{\partial_x W_{\pm}(x)}.
$$
The intersection points of the curves $\Sigma_+$ and $\Sigma_-$
$$\Sigma_+ \cap \Sigma_-,
$$
are the quantum vacua, where

$$
e^{{\partial_x }W(x)}= e^{\partial_x W_{+}(x)-\partial_x W_{-}(x)} =1.
$$
If the mass splittings within a flavor and the FI parameter $\zeta$ -- are non-vanishing, the vacua are isolated. If they are small, but non-zero, the position of the puncture on $C$ and the vacua coincide.
To each vacuum, and hence to each puncture on $C$, we associate an integer, $N_a$, $a=1, \ldots, M$, corresponding to the rank of the $3d$ gauge group there:
$$
\sum_{a=1}^{M} N_a = N.
$$

\subsection{5d Gauge Theory $\Sigma$}

In the 5d gauge theory ${\cal G}_{5d}$, like in 3d, the curve $C$ captures the classical, UV data.
Inserting a vertex operator in Liouville maps coupling the 5d theory to a pair of hypermultiplets
$H$, $H^{\dagger}$ in the fundamental and anti-fundamental representations, as well as raising the rank by $1$ (recall that ${\cal G}_{5d}$ is a $U(M)$ theory with $2M$ hypermultiplets).
The location of the puncture on $C$, $z_a$, and the choice of Liouville momentum  $\alpha_a$, get related to the 5d masses of two hypermultiplets. The integer $N_a$ associated to the vertex is mapped to the value of the Coulomb branch parameter $e_a.$

The Seiberg-Witten curve $\Sigma$ of ${\cal G}_{5d}$ can be viewed as a double cover of $C$,
$$
\Sigma \qquad \longrightarrow \qquad C.
$$
While $C$ captures the UV data, $\Sigma$ captures the vacuum structure of the theory.
$\Sigma$ arises by a geometric transition from the 3d curve $\Sigma_+ \cup \Sigma_-$,

$$\Sigma_+ \cup \Sigma_- \qquad \longrightarrow \qquad \Sigma.
$$
Before the transition, we have two sheets $\Sigma_+$ and $\Sigma_-$ that meet over $M$ points in the interior of $C$ where the plane curve $\Sigma_+ \cup \Sigma_- $ is singular. In the transition, the intersection points are replaced by cuts that smoothly join the two sheets. The size of the cut is measured by the period of the Seiberg-Witten 1-form $\lambda_{SW} = pdx$ around $1$-cycles $\gamma_a$ circling the cuts

\beq\label{period}
\oint_{\gamma_a} p dx = N_a \epsilon.
\eeq
The $M$ periods around $\gamma_a$ are normalizable, so they must be related to the $M$ Coulomb branch parameters.
These are, as we saw in section 4, equal to the ranks of the unbroken 3d gauge groups in the corresponding vacuua.

The emergence of the Seiberg-Witten curve $\Sigma$ from the 3d gauge theory can be seen explicitly in the large $N_a$-saddle point approximation to the partition function. The periods of the Seiberg-Witten differential compute the parameters $\log(e_a/e_b)$, $\log(e_a/f_b)$ that enter Nekrasov function. This parallels the story of geometric transitions in topological string theory \cite{GV, AKMV}, to which this reduces to at $q=t$. For $q\neq t$ similar examples were studied in \cite{Aganagic:2011mi}.

\subsection{Upstairs vs. downstairs}

There are different ways to split the contributions of chiral matter to $W(x)$ between $W_+(x)$ and $W_-(x)$; the choice of splitting affects both $\Sigma_+\cup\Sigma_-$ and $\Sigma$.
For a specific choice of $W_+$ and $W_-$, the we get
$$
\Sigma_+ \cup \Sigma_{-}: \qquad (e^p - e^{\partial_x W_+})(e^{-p} e^{\partial_x W_-}-1) =0,
$$
and corresponding to it
$$
\Sigma: \qquad e^{p}  + F(e^x)+  e^{-p} e^{\partial_x (W_++W_-)}=0.
$$
We can take all the chiral multiplets, coming from the denominator in \eqref{part}, to contribute to $W_+(x)$, and all the anti-chiral multiplets, from the numerator, to contribute to $W_-(x)$. Then,
$e^{\partial_x W_+} = \prod_{a=1}^M (1-e^{-x} e^{m_{a,+}}) $,  $e^{\partial_x W_-} = \prod_{a=1}^M (1-e^{-x} e^{m_{a,-}}) $, and $F$ is a polynomial of degree $M$ in $e^{-x}$; it is obtained by starting with $e^{\partial_x W_-} +e^{\partial_x W_+},$ both of which are polynomials of degree $M$ in $e^{-x}$, and deforming to a generic such polynomial.

Changing the splitting of $W$ into $W_+$ and $W_-$,
$$
W_{\pm}(x) \rightarrow W'_{\pm}(x)=W_{\pm} (x) + \Delta W(x)
$$
does not affect the theory of course; it leaves the downstairs curve $C$ the same, but it changes both $\Sigma_+\cup\Sigma_-$ and $\Sigma$.
By shifting a finite number of chiral multiplets form $W_+$ to $W_-$,
the curves one gets differ by a transformation that takes
$$
e^p \rightarrow e^{p'} = e^p e^{-\partial_x \Delta W}.
$$
where $e^{-\partial_x \Delta W}$ is a rational function of $z=e^x$.

The choice of splitting which affects the geometry does not affect the 3d gauge theory partition function ${\cal Z}_{{\cal G}_{3d}}$ or the 5d instanton partition function ${\cal Z}_{{\cal G}_{5d}}$, but it has a physicals counterpart. Both ${\cal Z}_{{\cal G}_{3d}}$ and ${\cal Z}_{{\cal G}_{5d}}$ compute indices counting BPS states. As we vary the parameters of the theory, what contributes to the index changes, leading to different ways of writing the index.  However the index itself is an analytic function of its parameters: all the different ways of writing the index are equivalent, related by analytic continuation.
In 3d, the ambiguity corresponds to trading a chiral multiplet $Q$, transforming in the fundamental representation, and of real mass $m$, with an anti-chiral multiplet $Q^{\dagger}$, transforming in the anti-fundamental representation, of mass $-m$. If $m>0$ BPS particles created by $Q$ contribute to the index, and those created by $Q^{\dagger}$ do not, as discussed in \cite{BDP}. If we then vary the mass from $m>0$ to $m<0$, it is $Q^{\dagger}$ that leads to particles that preserve the right supersymmetry to contribute to the index, while $Q$ does not. In the process of varying $m$ from positive to negative values, the 3d Chern-Simons term shifts by $1$, due to an anomaly \cite{Aharony:1997bx}.  In 5d, there is a completely analogous phenomenon. As we vary the 5d real mass $m$, a hypermultiplet $H$ of mass $m$ transforming in the fundamental representation gets traded with a hypermultiplet $H^{\dagger}$ transforming in the anti-fundamental representation and of real mass $-m$, at the expense of changing the 5d Chern-Simons level by $1$ \cite{Witten5dphases}. Once again, the 5d index before and after is the same, it is analytic in $m$, although we would write it in two different ways depending on whether $m>0$ or $m<0$. The 3d and 5d ambiguities match perfectly; the fact that they do is what makes the relation between the two theories possible.

The two different ways of writing the hypermultiplet contribution to the 5d index ${\cal Z}_{{\cal G}_{5d}}$ are described in
section 4, in \eqref{faf}. In the next subsection, we will discuss its 3d counterpart in more detail.

\subsection{Vertex Decomposition}

Dividing the curve $C$ into 3-punctured spheres divides the $\Sigma_+\cup \Sigma_-$ and  $\Sigma$ into vertices. From the 3d perspective, inserting a vertex corresponds to coupling the theory to a hypermultiplet in fundamental representation. There are 4 different ways of writing the hypermultiplet contribution to the partition function,

$$
\Phi_H(x)
$$
which lead to different ways of splitting $W(x)$ into $W_+(x)$ and $W_-(x)$ and
four possible choices of the vertex upstairs, on $\Sigma_+\cup \Sigma_-$. Since the geometric transition is local, this also leads to 4 different vertices for $\Sigma$,

$$V^{+-}, \qquad V^{-+}, \qquad V^{++}, \qquad V^{--},
$$
each coming from an ${\cal N}=4$ hypermultiplet in 3d. The choice of vertices does not affect ${\cal G}_{3d}$ or ${\cal G}_{5d}$, only their curves.

\begin{figure}[ht]
  \begin{center}
    \includegraphics[width=0.8\textwidth]{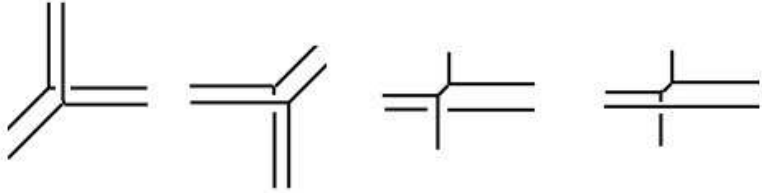}
  \end{center}
\vspace{-5ex}
  \caption{The four possible vertices, $V_{++}, V_{--}, V_{+-}$ and $V_{-+}$. 
 Poles in the graph are the vertical lines going up, where $e^p$ goes to infinity. Zeros are the vertical spikes running down, where $e^{p}$ goes to zero. In $V_{++}$ both $\Sigma_+$ and $\Sigma_-$ have a pole. In  $V_{+-}$, $\Sigma_+$ gets a pole and a zero.}\label{vertices}
\end{figure}
Written in the most obvious way,
\beq\label{basic2}
\Phi_{H}(x)= \prod_{1\leq i \leq N}  \prod_{n=0}^{\infty}{ 1 -q^{n} u \; e^{x_i-m_-}\over 1-q^n v \; e^{x_i-m_+}}.
\eeq
the hypermultiplet contributes to $W(x)$ as
$$W_H(x) = L(x-m_-) -L(x-m_+)
$$
where $L(x)$ is the classical dilogarithm. We get
$$V^{+-},
$$
by assigning $L(x-m_+)$ to $W_+$ and $L(x-m_-)$ to $W_-$. Then $\Sigma_{\pm}$ each get a pole at the position of the puncture in $C$. 

The hypermultiplet $H$ consists of a pair of chiral and anti-chiral multiplets $Q_+, \,Q_-^{\dagger}$ in the fundamental representation, which, as long as ${\rm Re}(x)>{\rm Re}(m_{\pm})$, both lead to BPS particles preserving the same supersymmetry, but have opposite statistics and opposite R-charge, leading to the denominator and the numerator of \eqref{basic2}.
Taking  ${\rm Re}(x)<{\rm Re}(m_{-})$ instead, one should think of $H$ as consisting of a pair of chiral multiplets $Q_+,{Q}_-$ which now have the same statistics but opposite charge, and same R-charge. This interpretation leads us to rewrite $\Phi_H$ as
$$
\Phi_{H}(x)= \prod_{1\leq i \leq N}  \prod_{n=0}^{\infty}{ 1\over (1 -q^{n} v \; e^{-x_i+m_-})(1-q^n v \; e^{x_i-m_+})} e^{{\sum_{i=1}^N}(x_i-m_-)^2/2\hbar}
$$
using quantum dilogarithm identities. The exponent quadratic in $x$ comes from the shift in the Chern-Simons level and FI terms as a BPS particle's mass goes through zero. This is just a rewriting of \eqref{basic}, but one which has the effect that we assign both $L(x-m_-)=(x-m_-)^2/2 - L(m_--x) $ and $L(x-m_+)$ to $W_+(x)$, and nothing to $W_-(x)$. This leads to the $V^{++}$ vertex. Then, $\Sigma_+$ gets a zero and a pole of $e^{p_+}$. Varying the masses further, both $Q_-^{\dagger}$ and ${ Q}_+^{\dagger} $ can become BPS, leading to the $V^{--}$ vertex, where both $L$'s contribute to $W_-$ instead, or to the $V^{-+}$ vertex, where $Q_+^{\dagger}$ and ${Q}_-$, both in anti-fundamental representation, are BPS, and $L(x-m_+)$ contributes to $W_-$ and $L(x-m_-)$ to $W_+$, see figure \ref{vertices}.

\begin{figure}[h]
  \begin{center}
    \includegraphics[width=0.7\textwidth]{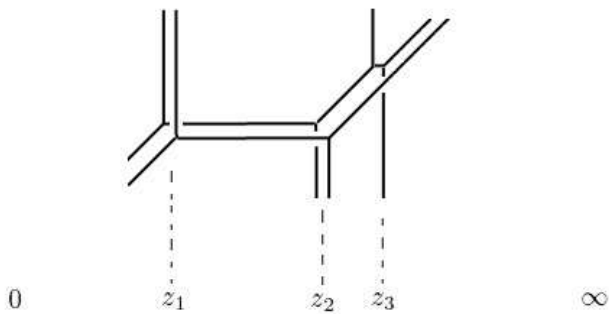}
  \end{center}
  \caption{Graph of a Riemann surface $\Sigma_{+}\cup {\Sigma_-}$ corresponding to $M=3$, and the 5-point conformal block. }\label{vert}
\end{figure}

When we replace $\Sigma_+\cup\Sigma_-$ by $\Sigma$, the different curves one gets lead to different ways of writing the Nekrasov function in \eqref{bN}, all differing by transformations \eqref{5dflip} and \eqref{iden}. They correspond to the same 5d gauge theory. The rational transformations that relate the different Seiberg-Witten curves $\Sigma$ affect the non-normalizable periods, but not the normalizable ones, see figure \ref{vert}.

\section{Discussion}

In this paper, we saw that there is a triality of correspondences between $q$-Liouville, the 3d gauge theory ${\cal G}_{3d}$ and the 5d gauge theory ${\cal G}_{5d}$. The Dotsenko-Fateev formulation of $q$-Liouville conformal blocks on a sphere with $M+2$ punctures is the partition function of ${\cal G}_{3d}$ $U(N)$ gauge theory, with an adjoint of mass $\epsilon$, and $M$ hypermultiplets. Evaluating the integrals by residues, we get a partition function of the 5d gauge theory ${\cal G}_{5d}$, at integral values of the Coulomb branch moduli. Arbitrary Coulomb branch moduli, which arise in the large $N$ limit of the 3d gauge theory, correspond to arbitrary $q$-Liouville conformal blocks, and arbitrary Coulomb branch moduli in ${\cal G}_{5d}$. What is striking about the triality is that it can be made essentially manifest, as we saw. A natural question is: Why? Why is this duality so easy to understand? There are two different ways to answer this, as we sketch in this section \cite{Simonstalk, toappear}. One is in terms of M-theory, and the other purely in terms of 5d gauge theory.  Finally, we discuss the relation to another Liouville/gauge theory correspondence, proposed in \cite{AGT}.

 \subsection{M-theory and Large $N$ transitions}
The gauge theory ${\cal G}_{3d}$ arises from $N$ M5 branes wrapping a three-cycle class in M-theory on a Calabi-Yau $X_{\Sigma_{\pm}}$.
The curves $\Sigma_{\pm}$, in the limit where they degenerate to graphs, are visible in the classical geometry of the Calabi-Yau $X$, as loci where $T^2$ fibration of $X_{\Sigma_{\pm}}$ degenerates, with 1-cycles of the $T^2$ shrinking over the edges in the graph.
Any three-cycle in this class ``stretches" between the two components of the curve (see \cite{AVG2} for details).  Taking $N$ M5 branes wrapping the homology cycle gives a $3d$ ${\cal N}=2$ $U(N)$ gauge theory. 
Almost everywhere, the graphs of $\Sigma_{\pm}$ run parallel to each other, and M5 branes wrap $S^2\times S^1$ (for $k=0$). The two graphs intersect at isolated points, where the effective Chern-Simons coupling is $\pm1$ and we get an $S^3$. We pick a vacuum, a distribution of $N$ M5 branes between the minimal $S^3$'s.
The partition function ${\cal Z}_{{\cal G}_{3d}}$ is the partition function of M-theory on $X_{\Sigma_{\pm}}\times M_{q,t}$, where the M5 branes wrap the 3-cycle class in $X_{\Sigma_{\pm}}$ and $M_q$ subspace of $M_{q,t}$; the complex plane in $M_{q,t} = ({\mathbb C} \times {\mathbb C} \times S^1)_{q,t}$ rotated by $t$ is transverse to the branes.

The gauge theory ${\cal G}_{5d}$ arises by compactifying M-theory on a Calabi-Yau $X_{\Sigma}$, obtained from $X_{\Sigma_\pm}$ by shrinking the minimal $S^3$'s wrapped by the M5 branes and replacing each with a ${\mathbb P}^1$.
The geometric transition from $X_{\Sigma_\pm}$ to $X_{\Sigma}$ is a transition from the Higgs branch of the 5d gauge theory ${\cal G}_{5d}$ to its Coulomb branch. The $N$ M5 branes are the vortices of ${\cal G}_{5d}$, that exist on the Higgs branch.

In general, the ${\cal G}_{3d}$ and ${\cal G}_{5d}$ theories are different; not even their dimensions match. This changes if we subject the 5d theory ${\cal G}_{5d}$ to the $\Omega$-background transverse to the vortex, with $\epsilon$ the parameter of the rotation. This effectively compactifies the 5d theory, preserving 2d Lorentz invariance and 4 supercharges \cite{NS2}, and allows one to turn on vortex flux, without introducing additional singularities \cite{Simonstalk, toappear}. On the ${\cal G}_{3d}$ side, the transverse $\Omega$-background with parameter $\epsilon$ gives mass to the adjoint scalar \cite{NS, NS1}.
Thus the 3d and the 5d theories can be dual to each other, at least at the level of BPS quantities. 

As explained in \cite{Simonstalk, toappear}, from M-theory perspective, the duality between ${\cal G}_{3d}$ and ${\cal G}_{5d}$ is a duality relating two different descriptions of M-theory on the same $singular$ Calabi-Yau $X$, with M5 brane 4-form flux turned on. In M-theory, the duality becomes manifest because the Calabi-Yau does not change. The Calabi-Yau $X$ sits at the transition point between  $X_{\Sigma_{\pm}}$ and $X_{\Sigma}$. Approaching $X$  from $X_{\Sigma_{\pm}}$ corresponds to a flow to the IR that sends the size of the $S^3$ to zero. Physics has to be non-singular there, as the 3d theory ${\cal G}_{3d}$, capturing the physics at the singularity, is massive. Approaching $X$ from $X_{\Sigma}$, the physics stays non-singular as well since the flux in $\Omega$-background gives masses to all the fields that would have gone massless there otherwise.

When $q=t$, Liouville field theory has an interpretation \cite{ DVt} as the topological string field theory of the topological B-model on the Seiberg-Witten curve $\Sigma$. The DF integrals have a physical interpretation, in the context of topological string, as theories on open topological B-branes before the transition, on ${\Sigma_+\cup \Sigma_-}$ \cite{DVt, IH}. For $q\neq t$ topological string has to be replaced by an M-theory index \cite{Dijkgraaf:2006um}, and large $N$ duality in M-theory \cite{Simonstalk, toappear}.
This naturally requires going to five dimensions. 
In this paper, we show that, in some special cases, the M-theory index can be traded for something simpler, a partition function of a supersymmetric gauge theory.

On a historical note: the BPS relations between the class of 3d and 5d theories studied in this paper were discussed going back to the work of \cite{DHT} and \cite{DH1,DH2}. The relation was revisited in \cite{DH1, DH2}, who studied the theories in the NS limit \cite{NS, NS1, NS2, NRSnew}, and interpreted the duality in  the context of integrable systems (see also \cite{Chen4, Chen3, Chen2, Chen1}).

\subsection{Instantons and Vortices}

It is natural to ask whether there is a direct way to understand the correspondence between summing over residues in the 3d gauge theory ${\cal G}_{3d}$ partition function and instanton counting in  ${\cal G}_{5d}$.

Expressing the partition function ${\cal Z}_{{\cal G}_{3d}}$ in terms of residues corresponds to trading integration with the Coulomb branch in ${\cal G}_{3d}$ to integration over the Higgs branch \cite{Bonelli:2011fq}. The poles of the integrand are the points on the Coulomb branch where  $M$ of the chiral multiplets in the theory become massless, and vortex solutions exist. The vortex charge $k = \int_{{\mathbb C}} {\rm Tr} F$ is detected by the coefficient of $e^{k \zeta}=q^{k \xi}$ in the partition function. For a pole labeled by an $M$-tuple of Young diagrams ${\vec R}$, $k=|{\vec R}|$, so a box in a Young diagram carries vortex charge $1$.
When we reinterpret the partition function in terms of ${\cal G}_{5d}$, $q^{\xi}$ measures instanton charge -- the same box in the Young diagram carries  ${\cal G}_{5d}$ instanton charge $1$. Thus, there has to be a one to one correspondence between torus fixed points in the moduli space of charge $k$ vortices in ${\cal G}_{3d}$ and the moduli space of charge $k$ instantons in ${\cal G}_{5d}$.

The theory on a charge $k$ vortex in the ${\cal G}_{3d}$ is a $0+1$ dimensional $U(k)$ gauge theory with four supercharges, an adjoint, as well as  $N$ chirals in the fundamental and $M-N$ chirals in the anti-fundamental representation. The contribution to ${\cal Z}_{3d}$ coming from charge $k$ vortices is computing the trace ${\rm Tr} (-1)^F $ of this quantum mechanics, with the twists coming from placing the 3d theory in $\Omega$-background $M_q$.  This can be evaluated by a 3d analogue of the localization formulas in \cite{Shadchin:2006yz, Bonelli:2011fq}; the latter were derived in the limit of vanishing circle size, when ${\cal G}_{3d}$ reduces to a 2d gauge theory.  One should compare the residue integrals that result to the well known localization formulas for ${\rm Tr}(-1)^F$ coming from quantum mechanics on the moduli space of charge $k$ instantons in ${\cal G}_{5d}$, when the latter is restricted to the locus \eqref{Coulomb}. It is easy to see that the formulas are essentially the same; the relation between them requires a single cancellation in the matter sector -- this is just rephrasing of the computation in section 4 in the language of \cite{Moore:1997dj, Losev:1997wp}, integrals. The relation between instanton counting in ${\cal G}_{5d}$ and vortex counting in ${\cal G}_{3d}$ is a statement that the instanton computation can be broken up into two steps: instead of going from 5d to 1d right away, as in the ADHM construction, we can first localize to 3d, then to 1d.

\subsection{Spectral Duality and AGT}

We found a triality of correspondences relating ${\cal G}_{3d}$,  ${\cal G}_{5d}$ and $q$-Liouville.  The relation between the 5d gauge theory and $q$-Liouville is reminiscent of AGT correspondence in 5d. There is the difference, though: ${\cal G}_{5d}$ is not the gauge theory AGT would have naturally associated to Liouville; rather it is its spectral dual ${\tilde {\cal G}}_{5d}$. M-theory on the Calabi-Yau $X_{\Sigma}$ has two different gauge theory descriptions. The description we have been using, corresponding to ${\cal G}_{5d}$, leads to 5d ${\cal N}=1$ theory with $U(M)$ gauge group and $2M$ fundamental hypermultiplets. The second description, ${\tilde {\cal G}}_{5d}$ is based on a $U(2)^{M-1}$ quiver theory. The fact that the two 5d gauge theories have the same M-theory origin is equivalent to the statement that they share the same Seiberg-Witten curve. In going from one to the other, one simply interprets the parameters of the theory differently, exchanging the role of the masses and the gauge couplings. As a result, the Nekrasov partition functions of the two theories look completely different. Under the assumption that the M-theory partition function on $X_{\Sigma}\times M_{q,t}$, constructed in \cite{ NOV} is well defined in the sense that it depends only on $X_{\Sigma}$, the K\"{a}hler moduli and $q$ and $t$, it equals the Nekrasov partition functions of ${\cal G}_{5d}$ and ${\tilde{\cal G}}_{5d}$ -- ${\cal Z}_{{\cal G}_{5d}}$ and
${\cal Z}_{{\tilde {\cal G}}_{5d}}$ are equal. The latter is simply the statement of geometric engineering. This, together with the results of our paper, then provides a proof of AGT correspondence in 5d, for conformal blocks on a sphere.

The original AGT correspondence arises in a limit where the $5d$ gauge theory ${\tilde{\cal G}}_{5d}$ becomes four dimensional. In the same limit, ${\tilde {\cal G}}_{3d}$ becomes a two dimensional gauge theory, and $q$-Liouville becomes ordinary Liouville. This also shows why proving AGT directly is hard -- in the same limit,
${\cal G}_{5d}$ and ${\cal G}_{3d}$ do not have any direct gauge theory interpretation. This said, the statement we can prove, namely the correspondence between ${q}$-Liouville conformal blocks and the partition function of ${\cal G}_{5d}$, is very much in the spirit of \cite{Gaiotto:2009we}. Namely, different ways of looking at the same Riemann surface make different aspects of the theory manifest. Looking at the curve $\Sigma$ as a double cover of the $M+2$ punctured sphere makes the connection to $A_1$ Toda CFT transparent. But, the relation of the 2d CFT to 5d gauge theory is only manifest if one views the same curve as an $M$-fold cover of the 4-punctured sphere.

\section*{Acknowledgments}
We thank the organizers and the participants of the 11th Simons Summer Workshop on Mathematics and Physics for a providing a stimulating environment.  We are especially grateful to Giulio Bonelli, Tudor Dimofte, Vasily Pestun, Joerg Teschner, Yuji Tachikawa, Alessandro Tanzini, Cumrun Vafa and Brian Wecht for valuable discussions. M.A, C.K. and S.S. thank Simons Center for Geometry for hospitality. C.K. is grateful to the Berkeley Center for Theoretical Physics and the Harvard University Theoretical High Energy Physics/String Theory group for hospitality. M.A. is also grateful to Lincoln Center for the Arts for hospitality. The research of M.A., N. H. and S. S is supported in part by the Berkeley Center for Theoretical Physics, by the National Science Foundation (award number 0855653), by the Institute for the Physics and Mathematics of the Universe, and by the US Department of Energy under Contract DE-AC02-05CH11231. The work of S. S is partly supported by grant RFBR 13-02-00478, grant for support of scientic schools Nsh-3349.2012.2 and government contract 8207. C.K. was partly supported by the INFN Research Project TV12.

\bibliographystyle{utcaps}	
\bibliography{myrefs}	
\end{document}